\documentclass[journal=apchd5,manuscript=letter]{achemso}

\usepackage[version=3]{mhchem} 
\usepackage[T1]{fontenc}       
\usepackage{epstopdf}
\usepackage{amsmath,amssymb,graphicx,bm}
\usepackage{graphicx}
\usepackage{color}
\usepackage{dcolumn}
\usepackage{bm}

\author{Alessandro Alberucci}
\email{alessandro.alberucci@tut.fi}
\affiliation{Optics Laboratory, Tampere University of Technology, FI-33101 Tampere, Finland}
\alsoaffiliation{N\textsl{oo}EL - Nonlinear Optics and OptoElectronics Lab, University ``Roma Tre'', IT-00146 Rome, Italy}

\author{Chandroth P. Jisha}
\affiliation{Centro de F\'{\i}sica do Porto, Faculdade de Ci\^encias, Universidade do Porto, PT-4169-007 Porto, Portugal}

\author{Lorenzo Marrucci}
\affiliation{Dipartimento di Fisica, Universit\`a di Napoli Federico II, IT-80100 Naples, Italy}
\alsoaffiliation{CNR-ISASI, Institute of Applied Science and Intelligent Systems, IT-80078 Pozzuoli (NA), Italy }

\author{Gaetano Assanto}
\affiliation{Optics Laboratory, Tampere University of Technology, FI-33101 Tampere, Finland}
\alsoaffiliation{N\textsl{oo}EL - Nonlinear Optics and OptoElectronics Lab, University ``Roma Tre'', IT-00146 Rome, Italy}
\alsoaffiliation{CNR-ISC, Institute for Complex Systems, IT-00185 Rome, Italy}

\title{Electromagnetic confinement via spin-orbit interaction in anisotropic dielectrics }

\keywords{Berry phase, waveguides, spin-orbit photonics, anisotropic media, structured beams, Kapitza effect}

\begin{document}



\begin{abstract}
We investigate electromagnetic propagation in uniaxial dielectrics with a transversely varying orientation of the optic axis, the latter staying orthogonal everywhere to the propagation direction. In such a geometry, the field experiences no refractive index gradients, yet it acquires a transversely-modulated Pancharatnam-Berry phase, that is, a geometric phase originating from a spin-orbit interaction. We show that the periodic evolution of the geometric phase versus propagation gives rise to a longitudinally-invariant effective potential. In certain configurations, this geometric phase can provide transverse confinement and waveguiding. The theoretical findings are tested and validated against numerical simulations of the complete Maxwell's equations. Our results introduce and illustrate the role of geometric phases on electromagnetic propagation over distances well exceeding the diffraction length, paving the way to a whole new family of guided waves and waveguides which do not rely on refractive index tailoring.
\end{abstract}

Several materials in nature feature an anisotropic - i.e., direction dependent -  electromagnetic response. Anisotropy is usually modeled by second-order tensors, in general coupling all the components of the electromagnetic field \cite{Kong:1990,Ghosh:2008,Khanikayev:2013,Bliokh:2014,Liu:2015}. 
The simplest case corresponds to non-magnetic anisotropic dielectrics, described by the constitutive equations $\bm{D}=\epsilon_0 \bm{\epsilon}\cdot \bm{E}$ and $\bm{B}=\mu_0 \bm{H}$, where $\bm{\epsilon}$ is the relative permittivity tensor, and $\epsilon_0$ and $\mu_0$ are the vacuum permittivity and permeability, respectively.  When the permittivity tensor is constant in space, for a given wavevector direction Maxwell's equations support two mutually orthogonal plane-wave eigensolutions with distinct refractive indices, i.e., exhibit birefringence. Orthogonality is no longer maintained when the optic axis is inhomogeneously rotated in the plane normal to the wavevector, yielding a point-wise accumulation of Pancharatnam-Berry phase (PBP) upon propagation. Only recently it has been recognized that the phase-front of a beam can be modified by means of the PBP \cite{Bomzon:2002,Marrucci:2006_1,Li:2015,Lin:2014,Tymchenko:2015}. \\ 
We refer to  monochromatic  waves (angular frequency $\omega$, time dependence $\propto e^{-i \omega t}$, vacuum wavelength $\lambda$ and wavenumber $k_0=2\pi/\lambda$) in uniaxial crystals (i.e., two out of the three eigenvalues of $\bm{\epsilon}$ coincide) with dielectric tensor $\bm{\epsilon}_D=\bm{\epsilon} =(\epsilon_\bot,0,0;0,\epsilon_\|,0;0,0,\epsilon_\bot)$ in the diagonal basis $x^\prime y^\prime z^\prime$, corresponding to the principal axes. The refractive indices for electric fields orthogonal or parallel to the optic axis are $n_\bot=\sqrt{\epsilon_\bot}$ or $n_\|=\sqrt{\epsilon_\|}$, respectively. We consider a uniaxial with a point-dependent rotation of $\bm{\epsilon}$ around the principal axis  $\hat{z}=\hat{z'}$, the latter coinciding with the direction of wave propagation. In such configuration the ordinary and extraordinary refractive indices remain always equal to $n_\bot$ and $n_\|$, respectively, and waves experience no spatial walk-off. Defining $\theta$ as the local angular rotation of $x^\prime y^\prime z^\prime$ with respect to the laboratory framework $xyz$ (Fig.~\ref{fig:sketch_geometry}) and using the circular polarization  basis $\hat{L}=(\hat{x}-i\hat{y})/\sqrt{2}$ (LCP, left-circular) and $\hat{R}=(\hat{x}+i\hat{y})/\sqrt{2}$ (RCP, right-circular), plane waves in this uniaxial dielectric evolve according to 
\begin{align}
&E_L(z) = e^{i\bar{n}k_0z}\left[\cos\left(\frac{\delta}{2}\right)E_L(0)-i\sin\left(\frac{\delta}{2}\right)e^{i2\theta}E_R(0)\right],  \nonumber\\
&E_R(z) = e^{i\bar{n}k_0z}\left[\cos\left(\frac{\delta}{2}\right)E_R(0)-i\sin\left(\frac{\delta}{2}\right)e^{-i2\theta}E_L(0)\right],
\label{eq:PW_R}
\end{align}
where $\delta(z)=k_0 z\Delta n$ is the retardation between ordinary and extraordinary components, $\Delta n = n_\| - n_\bot$ the birefringence and $\bar{n}=(n_\| + n_\bot)/2$ the average index of refraction \cite{Marrucci:2006}. Owing to birefringence, Eqs.~\eqref{eq:PW_R} describe a continuous power exchange between the two circular polarizations.\\ 
Let us consider a pure RCP at the input $z=0$ by setting $E_R(0)=1$ and $E_L(0)=0$. When $\delta=(2l+1)\pi$, with $l$ an integer, the RCP wave transforms into LCP: this change in polarization state is accompanied by a dynamic phase change $\Delta\phi_\mathrm{dyn}=(2l+1)\pi \bar{n}/\Delta n$ and  a phase shift $\Delta\phi_\mathrm{geo}= 2\theta$ of purely geometric origin, a manifestation of PBP \cite{Marrucci:2006}. Analogous dynamics occurs for a LCP input, now leading to an inverted geometric phase, i.e., $\Delta\phi_\mathrm{geo}= - 2\theta$. Equations~\eqref{eq:PW_R} show that geometric phases can modify electromagnetic wavefronts \cite{Marrucci:2006,Arbabi:2015,Bauer:2015,Zheng:2015,Li:2013}. However, since diffraction is neglected, Eqs.~\eqref{eq:PW_R} rigorously apply only to plane waves or wavepackets propagating in the anisotropic material for distances much smaller than the Rayleigh length. To study the propagation over longer distances, we must go beyond the plane-wave limit implicit in Eqs.~\eqref{eq:PW_R} and investigate the interplay between diffraction and the Pancharatnam-Berry geometric phase, as we will do in the following. \\
We note that, while the PBP is maximum in the planes where $\delta(z)=(2l+1)\pi$, it vanishes in the planes where  $\delta(z)=(2l)\pi$, and  takes intermediate values in between \cite{Slussarenko:2016}. Hence, in  homogeneous media, the PBP oscillates along  propagation, apparently without cumulative effects. As we reported recently \cite{Slussarenko:2016} and in analogy to quasi-phase matching in nonlinear optics, a periodic modulation of the optic axis along the propagation direction $z$ can yield a net cumulative PBP versus propagation, provided the period equals the beating length $\lambda/\Delta n$. Moreover, if the periodic modulation along the direction of propagation is associated with an inhomogeneous transverse distribution of the optic axis, waveguiding via the PBP can be achieved \cite{Slussarenko:2016}. \\
In this Letter we show that significant long-distance effects of the PBP on wave propagation can be obtained even in media perfectly invariant with respect to the propagation direction, that is, in the absence of longitudinal modulations of the optic axis distribution. We demonstrate, in particular, that the fast longitudinal modulation of the geometric phase gives rise to an effective $z$-invariant photonic potential through a Kapitza-like effect \cite{Alberucci:2013}. This  potential can support lateral confinement, paving the way to the realization of novel electromagnetic waveguides not based on refractive index changes, but on spin-orbit interactions between field polarization and wavefront. At variance with the case of longitudinally-modulated media \cite{Slussarenko:2016}, such PBP-confinement is \textit{independent} of the input polarization and it is immune from distributed reflections, a relevant feature in applications. \\
\section{Effective photonic potential}
\begin{figure} 
\includegraphics[width=0.5\textwidth]{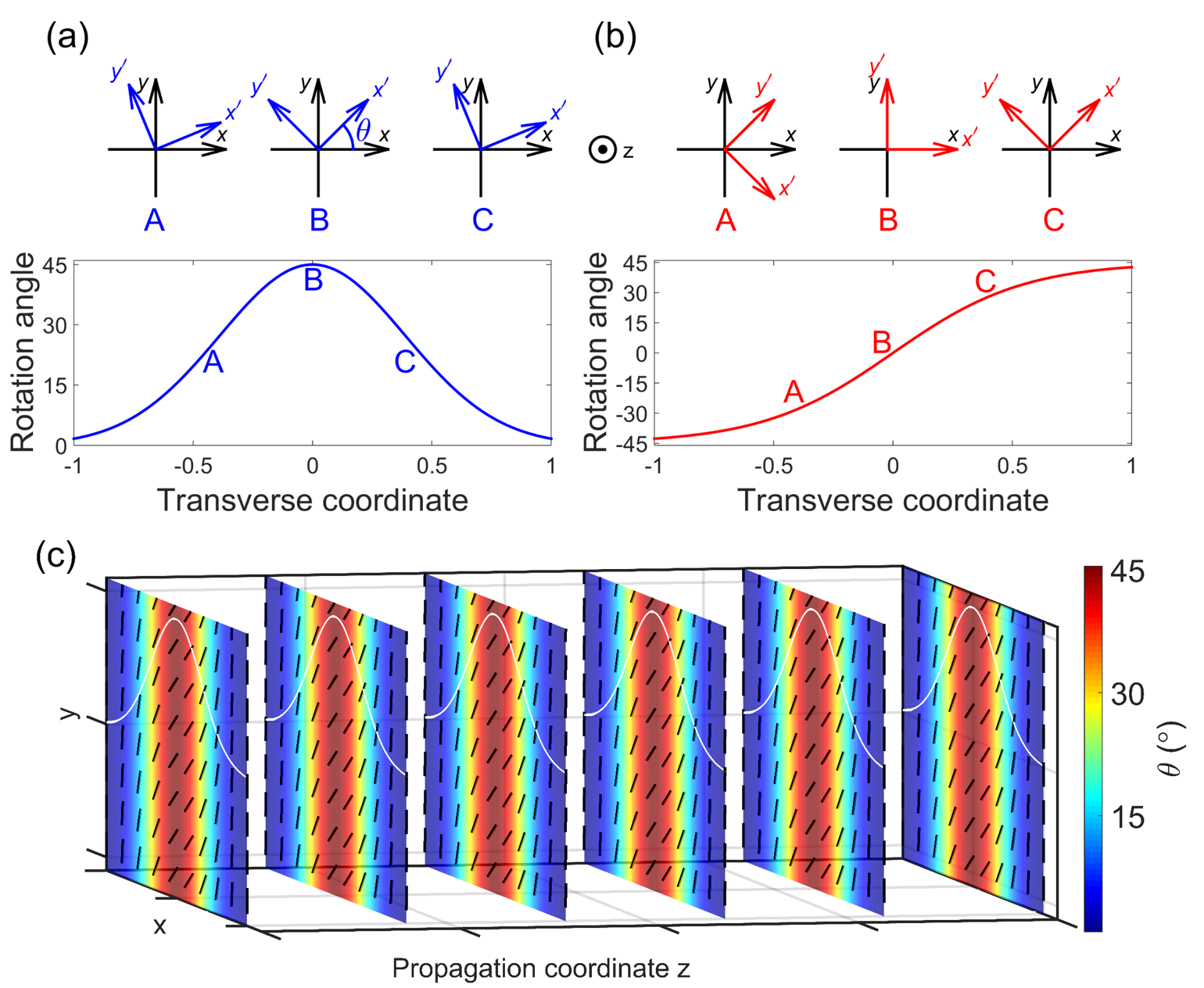}
\caption{\label{fig:sketch_geometry}  Material configuration in the transverse plane. Rotation angle $\theta$ of the principal axes $x^\prime y^\prime$ versus $xy$ when the distribution of $\theta$ is (a) Gaussian  or (b) hyperbolic tangent. The angle $\theta$ is positive when the rotation is counter-clockwise from the observer's point of view. The labeled points (A, B, C) in the graphs correspond to orientation of the principal axes as sketched above. In the plane-wave limit and for $\delta(z)=(2l+1)\pi$, distributions (a) and (b) yield a polarization-selective lens and a spin-dependent deflector based on the photonic spin-Hall effect, respectively. (c) 3D sketch of the optic axis distribution corresponding to the case plotted in panel (a). The structure is continuous and invariant along the propagation coordinate $z$ (only 6 slices are shown for the sake of clarity). The black rods correspond to the local optic axis. Solid white lines represent the corresponding profile of the rotation angle $\theta$ across $x$, also rendered by the superimposed color map. }
\end{figure}
For the sake of simplicity, we refer to a structure and fields varying only in the plane $xz$, a (1+1)D geometry resulting from setting $\partial_y=0$ in Maxwell's equations. The inhomogeneity consists of an $x-$dependent rotation of the principal axes of the uniaxial crystal (as sketched in Fig.~\ref{fig:sketch_geometry}). The principal eigenvalues $\epsilon_\bot$ and $\epsilon_\|$ of the relative permittivity tensor remain independent of the spatial position. Neglecting anisotropy in the diffraction operator \cite{Kwasny:2012}, in the laboratory framework $xyz$ Maxwell's equations for paraxial waves can be cast as 
\begin{equation} \label{eq:maxwell_equation}
 \nabla_{xz}^2 \left ( \begin{array} {c}
  E_x \\
  E_y \\
 \end{array} \right)   + k_0^2 \left ( \begin{array} {cc}
  \epsilon_{xx}(x) & \epsilon_{xy}(x) \\
  \epsilon_{yx}(x) & \epsilon_{yy}(x) \\
 \end{array} \right) \left ( \begin{array} {c}
  E_x \\
  E_y \\
 \end{array} \right)=0,
\end{equation}
with $\nabla_{xz}^2=\partial^2_x+\partial^2_z$. The paraxial approximation in Eq.~\eqref{eq:maxwell_equation} allows neglecting the longitudinal electric field, the latter relevant only for beam sizes comparable with or smaller than the wavelength \cite{Lax:1975}. Next, we rewrite Eq.~\eqref{eq:maxwell_equation} in a rotating framework locally aligned to the principal axes, introducing the 2D rotation operator $\bm{R}$ around the symmetry axis $\hat{z}$ 
\begin{equation}  \label{eq:rotation}
  \bm{R}(\theta) = \left ( \begin{array} {cc}
  \cos\theta & \sin\theta \\
  -\sin\theta & \cos\theta \\
 \end{array} \right),
\end{equation} 
with $\theta$ varying only across $x$, as in Fig.~\ref{fig:sketch_geometry}. Without loss of generality, we assume that for $\theta=0$ the optic axis is parallel to $\hat{y}$. The electric field in the rotated system is $ \bm{\varphi}=E_o(x,z) \widehat{x^\prime}(x) + E_e(x,z) \widehat{y^\prime}(x)$, with $E_o$ and $E_e$  the local ordinary and extraordinary components, respectively. After introducing the tensor $\bm{T}=\left(0,-1;1,0 \right)$ (proportional to Pauli's matrix $\bm{S}_2$, $\bm{T}=-i\bm{S}_2$), the application of Eq.~\eqref{eq:rotation} into Eqs.~\eqref{eq:maxwell_equation}  provides
\begin{align}
   \frac{\partial^2 \bm{\varphi}}{\partial z^2}  =
 - &  \frac{\partial^2 \bm{\varphi}}{\partial x^2} +  \left(\frac{d\theta}{d x} \right)^2 \bm{\varphi} - k_0^2  \bm{\epsilon}_D \cdot \bm{\varphi} \nonumber \\ &- \left( \frac{d^2 \theta}{d x^2}  \right) \bm{T}\cdot \bm{\varphi} - 2  \frac{d\theta}{d x} \bm{T} \cdot \frac{\partial \bm{\varphi}}{\partial  x}. \label{eq:maxwell_rotated_inho}
\end{align}
In Eq.~\eqref{eq:maxwell_rotated_inho} $\bm{\epsilon}_D$ is a diagonal matrix, $\bm{\epsilon}_D=(n^2_\bot,0;0,n^2_\|)$, constant in space due to the uniform distribution of $n_\bot$ and $n_\|$, and reduced to a 2$\times$2 tensor due to the transverse character of the electric field. The presence of terms depending on geometry (i.e., $\theta$ and its derivatives) resembles transformation optics \cite{Danner:2011}. All the terms containing $\bm{T}$ account for power exchange between ordinary and extraordinary components: owing to diffraction, a purely ordinary (extraordinary) wave at any given point is partially coupled into neighboring (transverse) regions with different $\theta$, thus yielding a mutual interaction between the two orthogonal polarizations, regardless of the reference system. Consistently, the size of such terms depends on the spatial derivatives of $\theta$. Such non-Abelian evolution and the absence of \textit{invariant modes} were  predicted earlier for light propagating in smoothly inhomogeneous anisotropic media \cite{Bliokh:2007}. \\
We apply the slowly varying envelope approximation through the transformation $E_o=e^{ik_0 n_\bot z}\psi_o$ and $E_e=e^{ik_0 n_\| z}\psi_e$, i.e., factoring out the dynamic phase responsible for polarization rotation versus propagation. For paraxial beams, Eq.~\eqref{eq:maxwell_rotated_inho} yields
\begin{align}
   2ik_0 n_\bot \frac{\partial {\psi}_o}{\partial z}  &=
 -   \frac{\partial^2 {\psi_o}}{\partial x^2} +  \left(\frac{d\theta}{d x} \right)^2 \psi_o  + \left( \frac{d^2 \theta}{d x^2}  \right)  \psi_e e^{ik_0\Delta n z} + 2  \frac{d\theta}{d x} \frac{\partial {\psi_e}}{\partial  x} e^{ik_0\Delta n z}, \label{eq:SVEA_ord} \\
2ik_0 n_\| \frac{\partial {\psi}_e}{\partial z}  &=
 -   \frac{\partial^2 \psi_e}{\partial x^2} +  \left(\frac{d\theta}{d x} \right)^2 \psi_e  - \left( \frac{d^2 \theta}{d x^2}  \right)  \psi_o e^{-ik_0\Delta n z}  - 2  \frac{d\theta}{d x} \frac{\partial {\psi_o}}{\partial  x} e^{-ik_0\Delta n z}. \label{eq:SVEA_ext}
	\end{align}
Equations~(\ref{eq:SVEA_ord}-\ref{eq:SVEA_ext}) indicate that the waves are not subject to any refractive index gradients, as the transverse phase modulation is only due to the point-wise rotation of the principal axes. For small birefringence $n_\bot\approx n_\|$, Eqs.~(\ref{eq:SVEA_ord}-\ref{eq:SVEA_ext}) resemble Pauli's equation for a charged particle of mass $m$, $ i\hbar\frac{\partial \bm{\psi}}{\partial t}=-\frac{\hbar^2}{2 m}\frac{\partial^2 \bm{\psi}}{\partial x^2} + U(x)\bm{I} \cdot \bm{\psi} + \bm{H}_{LS}(x,t) \cdot \bm{\psi} $, where $\bm{I}$ is the identity matrix \cite{Dirac:1999}. From the well-known analogy between 2D quantum mechanics and paraxial optics in the monochromatic regime, the propagation coordinate $z$ plays the role of time.  $\bm{H}_{LS}$ is a Hermitian matrix with zeroes in the main diagonal (i.e., a hollow matrix), accounting for  spin-orbit coupling \cite{Bliokh:2015}. In this analogy $\bm{\psi}$ is a two-component spinor with elements $\psi_o$ and $\psi_e$, respectively, $U(x)$ is a scalar potential acting equally on both components. $\bm{H}_{LS}(x,t)$ is proportional to $\bm{S}_2$ and equivalent to a time-dependent magnetic field normal to the particle spin and inducing spin-rotation (in our case a power exchange between extraordinary and ordinary components). \\
Following the similarity drawn above, the second terms on the RHS of Eqs.~(\ref{eq:SVEA_ord}-\ref{eq:SVEA_ext}) correspond to a \textsl{photonic potential} $V(x)=-k_0[n^2(x)-{n_j}^2]/(2{n_j}) \ (j=o,e)$  as it appears in the paraxial Helmholtz equation  $i\frac{\partial \psi}{\partial z}=-\frac{1}{2{n_j} k_0}\frac{\partial^2 \psi}{\partial x^2} + V(x) \psi$ in the presence of a refractive index distribution $n(x)$. In Eqs.~(\ref{eq:SVEA_ord}-\ref{eq:SVEA_ext}) the potential 
\begin{equation} 
  V(x)= \frac{1}{2n_j k_0}\left(\frac{d\theta}{dx} \right)^2 \label{eq:photonic_potential}
\end{equation}
arises from the transverse rotation of the dielectric tensor, which acts equally on both ordinary and extraordinary components. When the coupling terms on the RHS of Eqs.~(\ref{eq:SVEA_ord}-\ref{eq:SVEA_ext}) are negligible as compared to $V(x)$ (see Supporting Information), the latter can lead to transverse confinement and waveguiding with wavelength-independent mode profiles determined only by the spatial distribution of $\theta$, i.e., the geometric arrangement of the anisotropic dielectric. The wavelength independence of the mode profile stems from the fact that both the photonic potential given by Eq.~\eqref{eq:photonic_potential} and the effective mass in the particle-like model are inversely proportional to the vacuum wavenumber $k_0$. The effective potential Eq.~\eqref{eq:photonic_potential} originates from the periodic oscillation of the geometric phase $\Delta\phi_\mathrm{geo}$ along $z$, as the latter phase can be associated to a periodic potential $W(x) \sin\left( k_0 z \Delta n\right)$ acting on the wave. Owing to the Kapitza effect which stems from the transverse modulation of the effective kinetic energy \cite{Kapitza:1951,Alberucci:2013}, a $z$-invariant potential proportional to $\left(\frac{dW}{dx}\right)^2$ arises (see Supporting Information).\\
Wave propagation strongly depends on the symmetry of the distribution  $\theta (x)$. When $\theta$ is bell-shaped [Fig.~\ref{fig:sketch_geometry}(a)], the effective photonic potential has an inverted-W shape and supports leaky modes \cite{Alberucci:2013}. When $\theta (x)$ has an odd symmetry and $\frac{d\theta}{dx}\frac{d^2\theta}{dx^2}>0$ around $x=0$ [Fig.~\ref{fig:sketch_geometry}(b)], the photonic potential is maximum in the center: light gets repelled from the region around $x=0$ and no lateral confinement is expected. \\
\section{FDTD simulations}
\begin{figure}
\includegraphics[width=1\textwidth]{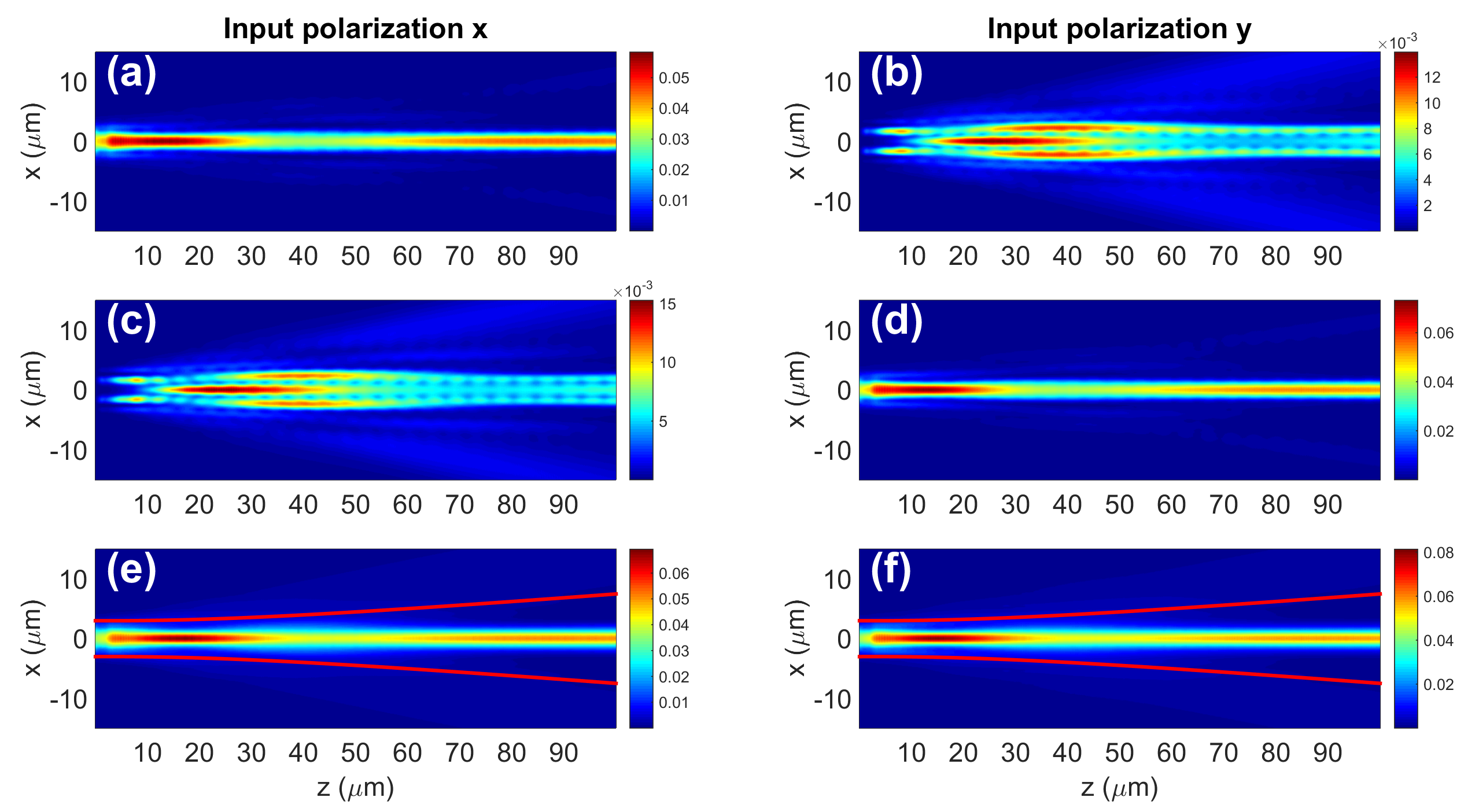}
\caption{\label{fig:FDTD_polarization} FDTD evolution in the plane $xz$ of an input wavepacket linearly polarized along either (left) $x$  or (right) $y$. Square of the electric field component along $x$ (upper row, a-b) and along $y$ (center row, c-d) averaged over a time period. (Bottom row, e-f) Time-averaged intensity. The red solid lines in (e,f) correspond to diffraction in a homogeneous sample ($\theta_0=0$) encompassing a Rayleigh length of $42\ \mu$m. Here $\theta_0=360^\circ$ and $w_\theta=5~\mu $m. In this case diffraction losses and the initial focusing are mainly ascribed to a mismatch between the input profile and the guided mode. }
\end{figure}
A first validation of the theory consists in verifying that the electromagnetic propagation is essentially independent from the input polarization. To this extent we assumed $\theta$ to be Gaussian by setting $\theta(x)=\theta_0 \exp\left({-\frac{x^2}{w_\theta^2}}\right)$. According to the theory, both polarizations sense the potential Eq.~\eqref{eq:photonic_potential}, thus should undergo confinement around $x=0$. Figure~\ref{fig:FDTD_polarization} shows the time-averaged wavepacket evolution for inputs linear polarized along either $x$ or $y$, respectively. The corresponding snapshots of the electric field can be found in the Supporting Information. The simulation results are in excellent agreement with Eq.~\eqref{eq:maxwell_equation} and confirm the theoretical predictions: the wave propagation depends negligibly on the input polarization, a result which is counter-intuitive  when considering the birefringence. The overall electromagnetic intensity is plotted in Fig.~\ref{fig:FDTD_polarization}(e-f): in agreement with Eq.~\eqref{eq:photonic_potential}, the optical wavepacket undergoes a marked lateral confinement as compared to a homogeneous sample, where the same input beam would diffract with a Rayleigh length of about 42 $\mu$m (red solid lines in the figure). Figure~\ref{fig:FDTD_polarization} also demonstrates that Eq.~\eqref{eq:maxwell_equation} satisfactorily approximates the complete set of Maxwell's equations for $\Delta n=0.2$. 
\begin{figure}
\includegraphics[width=1\textwidth]{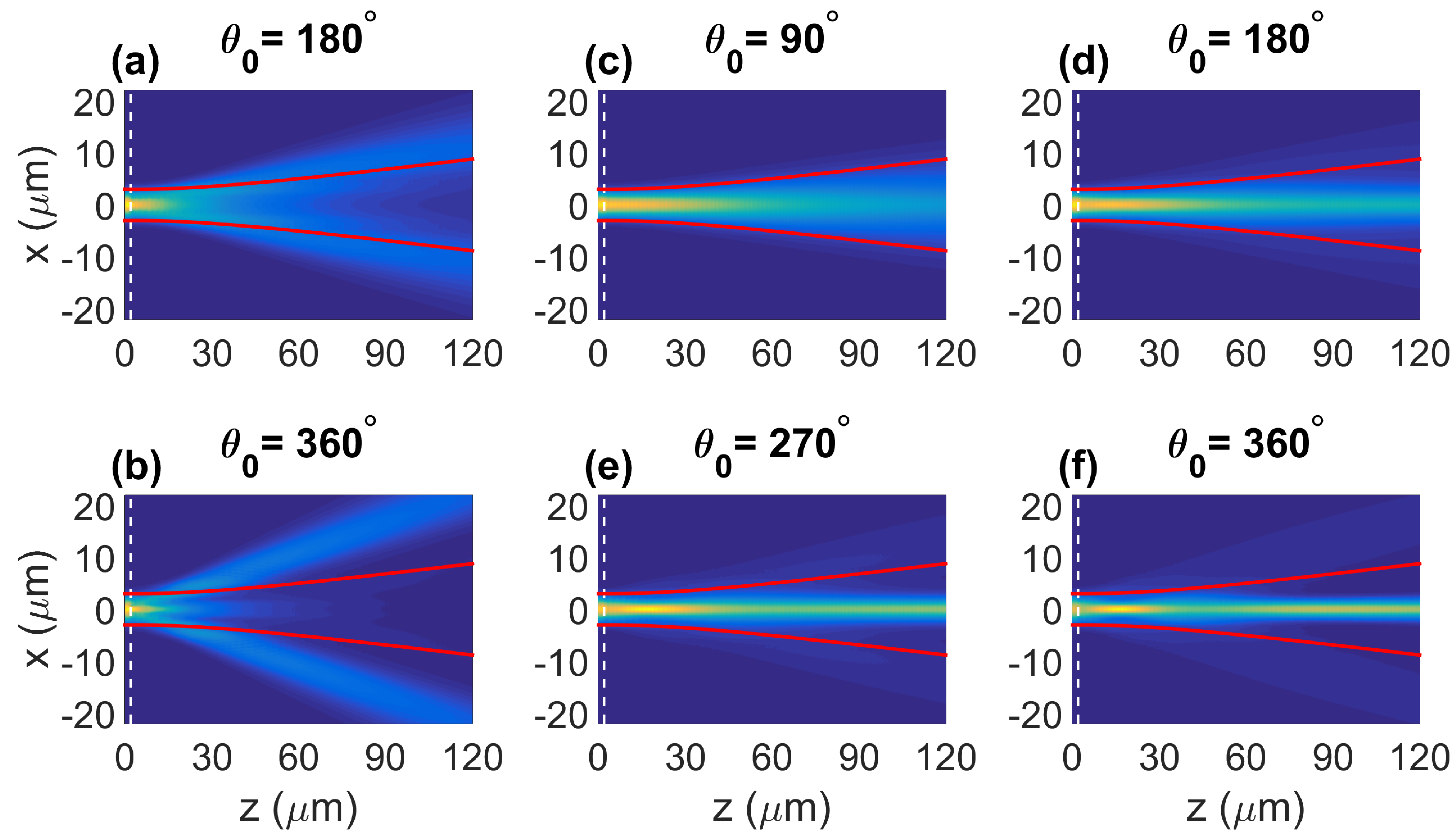}
\caption{\label{fig:FDTD_vs_theta_profile} FDTD simulations for  (a-b) the defocusing regime [$\theta(x)=\theta_0 \tanh{\left(\frac{x}{L_\theta}\right)}$ with $L_\theta=5~\mu $m] and (c-f) the waveguiding regime [$\theta(x)=\theta_0 \exp\left({-\frac{x^2}{w_\theta^2}}\right)$ with $w_\theta=5~\mu $m]. (a) $\theta_0=180^\circ$, (b) $\theta_0=360^\circ$, (c) $\theta_0=90^\circ$, (d) $\theta_0=180^\circ$, (e) $\theta_0=270^\circ$, and (f) $\theta_0=360^\circ$. The red solid lines show beam diffraction in a homogeneous medium, i.e., $\theta_0=0^\circ$. The white dashed lines correspond to the air/medium interface.}
\end{figure} 
Figure~\ref{fig:FDTD_vs_theta_profile} compares the propagation of a wavepacket in confining or repelling inhomogeneous structures, confirming the theoretical predictions. Confinement/repulsion become more effective for larger $\theta_0$, owing to a correspondingly stronger photonic potential. \\
\section{Waveguide design}
\begin{figure}
\includegraphics[width=0.5\textwidth]{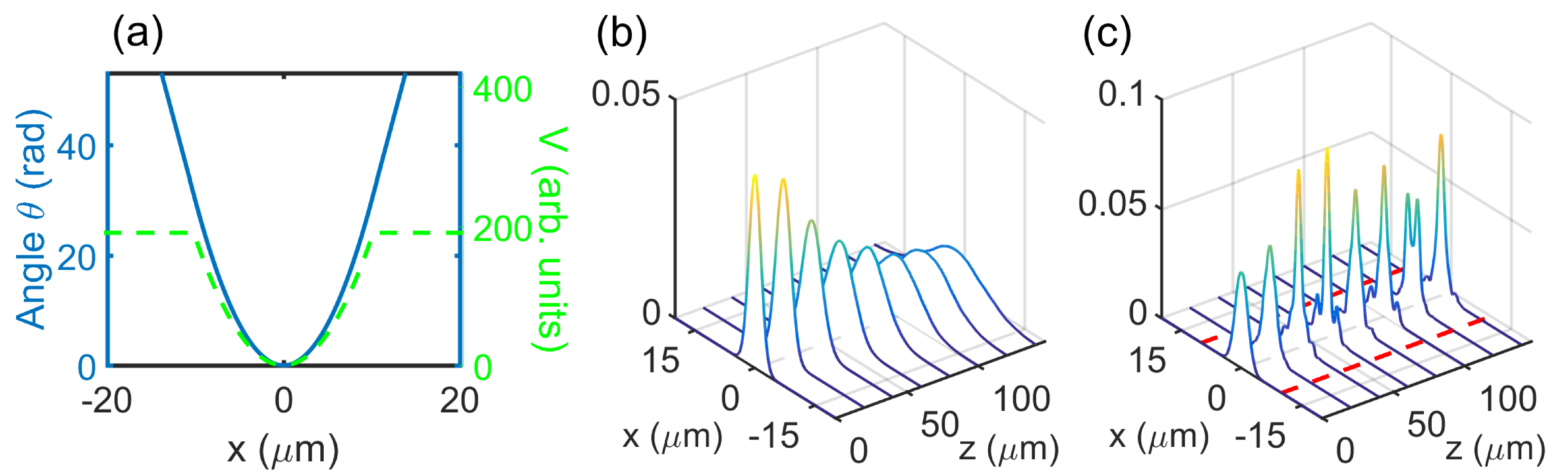}
\caption{\label{fig:guiding_parabolic} PBP guides featuring a V-shaped potential. (a) Rotation angle $\theta$ (blue solid line) and the corresponding potential $V$ from Eq.~\eqref{eq:photonic_potential} (green dashed line) for $x_0=10~\mu $m and $a=3\times 10^{11}~$m$^{-2}$. Time-averaged intensity extracted from FDTD simulations with (b) and without (c) the guiding structure. Red dashed lines mark the edges of the guide in $|x|=\pm x_0$.}
\end{figure}

Although the guides analyzed above  support  leaky modes, Eq.~\eqref{eq:photonic_potential} can be used to design a V-shaped photonic potential, thus canceling losses in the bulk region. For example, let us consider a rotation angle $\theta$ given by $ax^2$ for $|x|<x_0$, with a linear profile for $|x|>x_0$, its slope determined by the continuity of $\theta$ [see Fig.~\ref{fig:guiding_parabolic}(a)]. Equation~\eqref{eq:photonic_potential} provides a potential $V(x)=\frac{2a^2x^2}{2n_j k_0}\text{rect}_{2x_0}(x)$ supporting a finite number of guided modes [see Fig.~\ref{fig:guiding_parabolic}(a)]. The FDTD simulations in Fig.~\ref{fig:guiding_parabolic}(b-c) confirm that light gets confined, without coupling to the radiation modes in the bulk. \\
\section{Conclusions}
We investigated electromagnetic wave propagation in inhomogeneous uniaxials with a continuous rotation of the dielectric tensor around the wavevector direction. We found that the evolution of the wavepacket profile is governed by the geometric phase. A beam of arbitrary wavelength can be either focused or defocused according to the spatial dependence of the optic axis rotation in the transverse plane, leading to lateral trapping for a bell-shaped distribution. Quite counter-intuitively, in the absence of walk-off an inhomogeneously twisted anisotropic medium can provide an overall isotropic response. The results apply to all frequencies in the electromagnetic spectrum and were validated against numerical simulations. They could find applications towards a brand new class of waveguides based on geometric phases. Potential systems for the experimental demonstration include liquid crystals \cite{Kobashi:2016}, metastructures \cite{Khorasaninejad:2016,Jahani:2016}, laser-nanostructured glasses \cite{Beresna:2011}. Future developments include investigating the connections with  gauge optics \cite{Lin:2014_1,Liu:2015} and the interplay between PBP and spin redirection Berry phase \cite{Bliokh:2008}. 

\begin{acknowledgement}
A.A. and G.A. thank the Academy of Finland for support through the Finland Distinguished Professor grant no. 282858. C.P.J.  acknowledges Funda\c{c}\~{a}o para a Ci\^{e}ncia e a Tecnologia, POPH-QREN and FSE (FCT, Portugal) for the fellowship SFRH/BPD/77524/2011. L.M. is grateful to the European Research Council (ERC) for support under grant No. 694683, PHOSPhOR.
\end{acknowledgement}

\section{Methods}
For the numerical simulations we employed the open-source finite-difference time-domain (FDTD) code MEEP \cite{Oskooi:2010} and continuous-wave excitation, corresponding to $\lambda=1~\mu $m. The used frequency was chosen in the optical spectrum due to its relevance for applications, although our findings are valid regardless of the wavelength. The source was a Gaussian-shaped dipole ensemble, $3~\mu $m wide across $x$, infinitesimally narrow along $z$ and centered in $x=z=0$. The uniaxial medium starts at $z=2~\mu $m, with refractive indices $n_\bot=1.5$ and $n_\|=1.7$, respectively, corresponding to standard nematic liquid crystals, where the optic axis can be rotated locally \cite{Marrucci:2006}.

\begin{suppinfo}

Supporting Information Available: 
\begin{itemize}
  \item PDF file containing the derivation of the eigenvalue problem determining the quasi-mode, the interpretation of the effective potential as a Kapitza effect stemming from the Pancharatnam-Berry phase and the snapshot of the optical field computed via the FDTD simulations. 
\end{itemize}
 This material is available free of charge via the Internet at http://pubs.acs.org.
\end{suppinfo}

\providecommand{\latin}[1]{#1}
\makeatletter
\providecommand{\doi}
  {\begingroup\let\do\@makeother\dospecials
  \catcode`\{=1 \catcode`\}=2\doi@aux}
\providecommand{\doi@aux}[1]{\endgroup\texttt{#1}}
\makeatother
\providecommand*\mcitethebibliography{\thebibliography}
\csname @ifundefined\endcsname{endmcitethebibliography}
  {\let\endmcitethebibliography\endthebibliography}{}

\end{document}